\documentclass[aps,prb,reprint,amsmath,amssymb]{revtex4}
\usepackage{graphicx}
\usepackage{dcolumn}
\usepackage{bm}
\usepackage{float}
\usepackage{xcolor}
\usepackage{hyperref}
\hypersetup{colorlinks=true, citecolor=blue, urlcolor=blue, linkcolor=blue}
\begin{document}
\title{Full description of dipole orientation in organic light-emitting diodes}
\author{Lingjie Fan$^{1,2}$}
\thanks{These authors contributed equally to this work.}
\author{Tongyu Li$^{1,2}$}
\thanks{These authors contributed equally to this work.}
\author{Jiao Chu$^{1}$}
\author{Maoxiong Zhao$^{1,2}$}
\author{Tangyao Shen$^{1,2}$}
\author{Minjia Zheng$^{1,2}$}
\author{Fang Guan$^{4}$}
\author{Haiwei Yin$^{2}$}
\author{Lei Shi$^{1,2,3,4}$}
\email{lshi@fudan.edu.cn}
\author{Jian Zi$^{1,2,3,4}$}
\email{jzi@fudan.edu.cn}

\affiliation{$^{1}$Department of Physics, Key Laboratory of Micro- and Nano-Photonic Structures (Ministry of Education), and State Key Laboratory of Surface Physics, Fudan University, Shanghai 200433, China}
\affiliation{$^{2}$Shanghai Engineering Research Center of Optical Metrology for Nano-fabrication (SERCOM), Shanghai 200433, China}
\affiliation{$^{3}$Collaborative Innovation Center of Advanced Microstructures, Nanjing University, Nanjing 210093, China}
\affiliation{$^{4}$Institute for Nanoelectronic devices and Quantum computing, Fudan University, Shanghai 200438, China}
\date{\today}

\begin{abstract}
Considerable progress has been made in organic light-emitting diodes (OLEDs) to achieve high external quantum efficiency, among which the dipole orientation of OLED emitters has a remarkable effect. In most cases, external quantum efficiency of the OLED emitter is theoretically predicted with only one orientation factor to match with corresponding experiments. Here, we develop a distribution theory with three independent parameters to fully describe the relationship between dipole orientations and power densities. Furthermore, we propose an optimal experiment configuration for measuring all the distribution parameters. Measuring the unpolarized spectrum can dig more information of dipole orientation distributions with a rather simple way. Our theory provides an universal plot of the OLED dipole orientation, paving the way for designing more complicated OLED structures.
\end{abstract}
\maketitle

\section{Introduction}
Since the first reports of OLEDs in 1987~\cite{Tang1987Org}, its efficiency has been improved through finding novel phosphorescent materials~\cite{Naka2014High,Kim2005High,Kim2018High,Baldo1998Highly,Baldo1999Very,Helan2011Chlor,Kim2018Origin,Costa2010Dumb}, optimizing the thickness of each layer in OLED stacks~\cite{Lin2006Enhan,Flae2009In,Nowy2009light}, and so on. In recent years, the emitting dipole orientation has also been drawing significant attention for enhancing light extraction from OLED stacks~\cite{Fris2010Deter,Brut2013Device,Flae2010Mea,Mor2016Dep,Kom2017Dip,Cho2018Non,Has2018Well}.

The theory of dipole radiation in OLEDs originated from applying the theory of electrical dipoles near an interface to the problem of molecules fluorescing near a surface~\cite{Chan1974Life,Luk1977Light,Chan2007Mole,Luk1980Theory,Luk1981Light,Ford1984Elec,Netys1998Simu,Bar1998Fluo}. Related researches have been proposed to enhance the external quantum efficiency~\cite{Schm2017Emit}, in which the description of the dipole orientation distribution is applied. In previous description, A dipole ensemble is decomposed into vertical dipole and horizontal dipole. The power radiated by the dipole ensemble is formed by weighting the power radiated by vertical dipole and horizontal dipole in a proportion. The scale factor, the ratio of vertical dipole and horizontal dipole, can be regarded as a parameter describing the vertical orientation distribution of the dipole ensemble.

Recently along with the OLED manufacturing progress increasingly, it is pointed out that a possible research direction of OLEDs in the future is to take advantage of the dipole ensemble with non-uniform horizontal orientation distribution~\cite{Yoko2011Mole}. Some studies point out that the carrier mobility in films using the dipole ensemble perfect aligning in one direction is much higher than that of films using the uniform dipole ensemble~\cite{Sun2004Elas,Amaya2009Aniso}. At the same time, other researchers show that OLEDs in which dipoles align in one direction can achieve the emission of linear polarized light. Then, let the linear polarized light pass through a quarter-wave plate formed by the liquid crystal to realize the application of OLEDs directly emitting orthogonal circular polarized light~\cite{Baek2019Simu}. Emitting dipole orientation is of great significance for the external quantum efficiency, as well as the polarization. However, the previous works have so far only considered the dipole ensemble with non-uniform vertical orientation distribution. Lack of the description of the horizontal component, the performance of the OLED devices adjusted by only one vertical orientation factor is limited by the low dimension of regulation, e.g. the polarization performance of OLED devices is greatly influenced by the horizontal component, which is not contained in previous theory. With the further study of OLEDs and more precise control of the horizontal dipole orientation, there is no doubt that a theory fully describing the relationship between dipole orientation distribution and power density is needed.

Here, we develop a theory that fully describes the relationship between the orientation distribution of a dipole ensemble and its power density. In contrast to the decomposition method used in previous theory, we start from a dipole and consider an arbitrary distribution function, expressed as a Fourier series, to extend the power density of a dipole to the power density of a dipole ensemble. Theoretically, it is strictly proved that only three distribution parameters are needed to fully describe the effect of orientation distribution of a dipole ensemble on its power density. Two parameters describe the vertical and horizontal orientation distribution, and one parameter describes the coupling between the vertical and horizontal distribution. Finally, by using optical simulation, we design an experimentally feasible scheme for measuring these three distribution parameters, and present different spectra corresponding to different distribution parameters for the test structure.

\begin{figure}[t]
\centering
\includegraphics[scale=1]{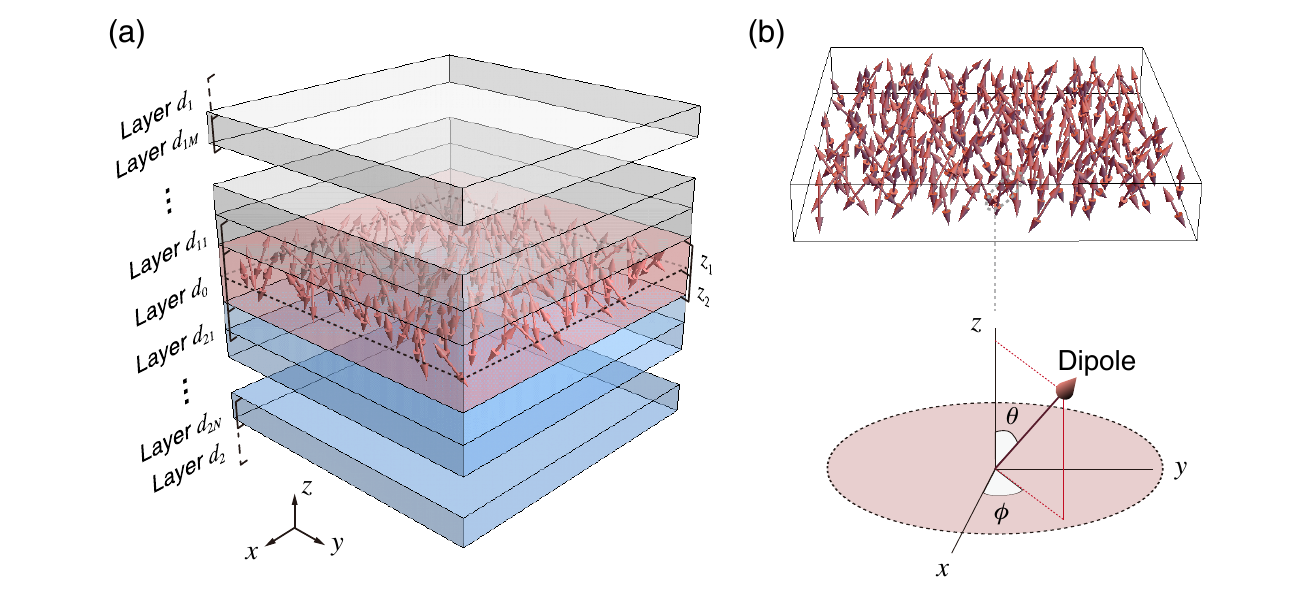}
\caption{\label{fig1}
(a)Sketch of OLED stacks. (b) Upper panel: dipole ensemble in the 0 layer. Lower panel: the orientation of a dipole.}
\end{figure}

\section{The power density of a dipole in the stacks}
The theoretical model of the OLED stacks is shown in Fig.~\ref{fig1}(a). The emitting layer with an index of refraction $n_{0}$ and thickness $d_{0}$ is located between two stacks of layers. The intermediate layers $j1,~j2,~\cdots$ have indices of refraction $n_{j1},~n_{j2},~\cdots$ and thicknesses $d_{j1},~d_{j2},~\cdots$. The half-infinite spaces are labeled $j~\left(j=1,~2\right)$. The dipole ensemble in layer 0 is located at a distance $z_{j}$ from the interface at the $j$ side of the layer. The OLED stacks, therefore, can be simplified to the multi-layer films. When there is only one electric dipole in the OLED stacks, its orientation is shown in Fig.~\ref{fig1}(b) lower panel.

For the multi-layer model shown in Fig.~\ref{fig1}(a), the energy reflection and transmission coefficients are given by
\begin{subequations}\label{eq1}
\begin{align}
R^{s,p}_{j}=\left|r^{s,p}_{j}\right|^{2} &\qquad \textmd{for all cases}, \\
T^{s}_{j}=\left|t^{s}_{j}\right|^{2} \frac{k_{z,j}}{\left|k_{z,0}\right|} &\qquad \textmd{for Im}\left(k_{z,j}\right)=0, \\
T^{p}_{j}=\left|t^{p}_{j}\right|^{2} \frac{n^{2}_{0}}{n^{2}_{j}} \frac{k_{z,j}}{\left|k_{z,0}\right|} &\qquad \textmd{for Im}\left(k_{z,j}\right)=0, \\
T^{s,p}_{j}=0 &\qquad \textmd{for Im}\left(k_{z,j}\right)\neq 0.
\end{align}
\end{subequations}
$r^{s,p}_{j}$, $t^{s,p}_{j}$ are Fresnel reflection and transmission coefficients, which could be calculated through recursive matrix algorithms~\cite{Katsidis:02,Li:96}. The energy coefficients and Fresnel coefficients with subscripts $j$ represent the reflection and transmission coefficients from the organic emitting layer (layer 0) to the half-spaces $j~(j=1,~2)$. $k_{\parallel}$ and $k_{z,j}$ are the radial and $z$ component of the wave vector $k_{j}$ in the half-spaces $j$.
With the use of the superposition of plane waves, the power $W_{j}$ radiated to the half-spaces $j$ can be written as an integral:
\begin{equation}\label{eq2}
W_{j}=\int^{+ \infty}_{0}K_{j}\left(k_{\parallel},\theta,\phi\right)~dk^{2}_{\parallel},
\end{equation}
where $K_{j}\left(k_{\parallel},\theta,\phi\right)$ is the power density per unit $dk^{2}_{\parallel}$. Based on the polarity, we can separate the power density $K_{j}\left(k_{\parallel},\theta,\phi\right)$ into $s$-polarized power density $K^{s}_{j}\left(k_{\parallel},\theta,\phi\right)$ and $p$-polarized power density $K^{p}_{j}\left(k_{\parallel},\theta,\phi\right)$:
\begin{equation}\label{eq3}
K_{j}\left(k_{\parallel},\theta,\phi\right)=K^{s}_{j}\left(k_{\parallel},\theta,\phi\right)+K^{p}_{j}\left(k_{\parallel},\theta,\phi\right),
\end{equation}
where polar angle $\theta$ and azimuth angle $\phi$ are shown in Fig.~\ref{fig1}(b) lower panel. For an electric dipole in the emitting layer, the $s$-polarized and $p$-polarized power densities radiated to the half-spaces $j$ in $x-z$ plane per unit $d k_{\parallel}^{2}$ read~\cite{Luk1981Light,Netys1998Simu}
\begin{subequations}\label{eq4}
\begin{align}
\begin{split}
K^{s}_{j}\left(k_{\parallel},\theta,\phi\right) &= \frac{3}{8}\frac{1}{k_{0}k_{z,0}}\frac{(1+a^{s}_{3-j})\overline{(1+a^{s}_{3-j})}}{\left|1-a^{s}\right|^{2}}T^{s}_{j}\sin^{2}\theta\sin^{2}\phi \\
&=\frac{3}{8}\frac{1}{k_{0}k_{z,0}}\frac{\left|1+a^{s}_{3-j}\right|^{2}}{\left|1-a^{s}\right|^{2}}T^{s}_{j}\sin^{2}\theta\sin^{2}\phi,
\end{split}\\
\begin{split}
K^{p}_{j}\left(k_{\parallel},\theta,\phi\right) &=
\frac{3}{8}\frac{k^{2}_{\parallel}}{k^{3}_{0}k_{z,0}}\frac{(1+a^{p}_{3-j})\overline{(1+a^{p}_{3-j})}}{\left|1-a^{p}\right|^{2}}T^{p}_{j}\cos^{2}\theta \\
 &+\frac{3}{8}\frac{k_{z,0}}{k^{3}_{0}}\frac{(1-a^{p}_{3-j})\overline{(1-a^{p}_{3-j})}}{\left|1-a^{p}\right|^{2}}T^{p}_{j}\sin^{2}\theta\cos^{2}\phi
\\
 &-\frac{3}{8}\frac{k_{\parallel}}{k^{3}_{0}}\frac{(1+a^{p}_{3-j})\overline{(1-a^{p}_{3-j})}+\overline{(1+a^{p}_{3-j})}(1-a^{p}_{3-j})}{\left|1-a^{p}\right|^{2}}T^{p}_{j}\sin
2\theta \cos \phi,
\end{split}
\end{align}
\end{subequations}
where $a^{s,p}_{j}$ and $a^{s,p}_{3-j}$ are given by
\begin{subequations}\label{eq5}
\begin{align}
a^{s,p}_{j} &= r^{s,p}_{j}\exp\left(2ik_{z,0}z_{j}\right), \\
a^{s,p}_{3-j} &= r^{s,p}_{3-j}\exp\left(2ik_{z,0}z_{3-j}\right),
\end{align}
\end{subequations}
and $a^{s,p}$ is the product of $a^{s,p}_{j}$ and $a^{s,p}_{3-j}$
\begin{equation}\label{eq6}
\begin{split}
a^{s,p}&=a^{s,p}_{j}a^{s,p}_{3-j}\\
~&=r^{s,p}_{j}r^{s,p}_{3-j}\exp\left(2ik_{z,0}d_{0}\right).
\end{split}
\end{equation}
$\overline{(1-a^{p}_{3-j})}$ and $\overline{(1+a^{p}_{3-j})}$ represent the conjugate of $(1-a^{p}_{3-j})$ and $(1+a^{p}_{3-j})$, respectively.

According to the specific form of power densities given by Eq.~\eqref{eq4}, $p$-polarized power density $K^{p}_{j}\left(k_{\parallel},\theta,\phi\right)$ can be further simplified and separated into three different parts $K^{p1}_{j}\left(k_{\parallel},\theta,\phi\right)$, $K^{p2}_{j}\left(k_{\parallel},\theta,\phi\right)$ and $K^{p3}_{j}\left(k_{\parallel},\theta,\phi\right)$:
\begin{equation}\label{eq7}
K^{p}_{j}\left(k_{\parallel},\theta,\phi\right)=K^{p1}_{j}\left(k_{\parallel},\theta,\phi\right)+K^{p2}_{j}\left(k_{\parallel},\theta,\phi\right)+K^{p3}_{j}\left(k_{\parallel},\theta,\phi\right),
\end{equation}
with
\begin{subequations}\label{eq8}
\begin{align}
K^{p1}_{j}\left(k_{\parallel},\theta,\phi\right) &=
\frac{3}{8}\frac{k^{2}_{\parallel}}{k^{3}_{0}k_{z,0}}\frac{\left|1+a^{p}_{3-j}\right|^{2}}{\left|1-a^{p}\right|^{2}}T^{p}_{j}\cos^{2}\theta, \\
K^{p2}_{j}\left(k_{\parallel},\theta,\phi\right) &=
\frac{3}{8}\frac{k_{z,0}}{k^{3}_{0}}\frac{\left|1-a^{p}_{3-j}\right|^{2}}{\left|1-a^{p}\right|^{2}}T^{p}_{j}\sin^{2}\theta\cos^{2}\phi, \\
K^{p3}_{j}\left(k_{\parallel},\theta,\phi\right) &= -\frac{3}{8}\frac{k_{\parallel}}{k^{3}_{0}}\frac{1}{\left|1-a^{p}\right|^{2}}T^{p}_{3-j}T^{p}_{j}\sin 2\theta \cos \phi.
\end{align}
\end{subequations}

For an emission angle $\alpha$, the angle between $z$ axis and the detector (observer) in half-space $j$, the power density $K^{s,p}_{j}\left(k_{\parallel},\theta,\phi\right)$ per unit $d k_{\parallel}^{2}$ can be transformed to the power density per solid angle $P^{s,p}_{j}\left(\alpha,\theta,\phi\right)$, given by
\begin{equation}\label{eq9}
P^{s,p}_{j}\left(\alpha,\theta,\phi\right)=\frac{k_{j}^{2}\cos\alpha}{\pi} K^{s,p}_{j}(k_{\parallel},\theta,\phi).
\end{equation}

\section{Full description of orientation distribution}
For a dipole ensemble with arbitrary orientation distribution in the OLED stacks, we use a distribution function $F\left(\theta,\phi\right)$ to describe its orientation distribution. Multiplied with distribution function and then integrated over the sphere with the polar angle $\theta$ and azimuth angle $\phi$, the power density can be used to describe a dipole ensemble:
\begin{equation}\label{eq10}
\widetilde{P}\left(\alpha,a,e,c\right)=\int^{2 \pi}_{0}\int^{\pi}_{0} P\left(\alpha,\theta,\phi\right)F\left(\theta,\phi\right) ~d\theta~d\phi,
\end{equation}
where $\widetilde{P}\left(\alpha, a,e,c\right)$ is the power density of a dipole ensemble adjusted by three distribution parameters. The distribution parameters $a$, $e$, and $c$ will be given through further derivation. The factor $\sin\theta$ in the spherical integral is included in the distribution function $F\left(\theta,\phi\right)$ for simplicity. Thus, the distribution function of a dipole ensemble with three-dimensional isotropical random orientation distribution is given by:
\begin{equation}\label{eq11}
F\left(\theta,\phi\right)=\frac{1}{4\pi}\times 1 \times \sin\theta.
\end{equation}

By using the Fourier series expansion, we can expand the distribution function $F\left(\theta,\phi\right)$ into Fourier series
\begin{equation}\label{eq12}
\begin{split}
F\left(\theta,\phi\right) &=\sum_{m,n=1}^{\infty}\lambda_{m,n}(a_{m,n} \cos \left(2m\theta\right)\cos \left(n\phi\right)+b_{m,n}\cos\left(2m\theta\right)\sin\left(n\phi\right)\\
~&+c_{m,n}\sin\left(2m\theta\right)\cos\left(n\phi\right)+d_{m,n}\sin\left(2m\theta\right)\sin\left(n\phi\right)),
\end{split}
\end{equation}
with
\begin{subequations}\label{eq13}
\begin{align}
\lambda_{m,n}=\frac{1}{4} &\qquad \textmd{for}~m=0,~n=0,\\
\lambda_{m,n}=\frac{1}{2} &\qquad \textmd{for}~m=0,~n\neq0~\textmd{or}~m\neq0,n=0,\\
\lambda_{m,n}=1 &\qquad \textmd{for}~m\neq0,~n\neq0.
\end{align}
\end{subequations}
and coefficients $a_{m,n},~b_{m,n},~c_{m,n},~d_{m,n}$ are given by
\begin{subequations}\label{eq14}
\begin{align}
a_{m,n} &= \frac{2}{\pi^{2}} \int_{0}^{2\pi} \int_{0}^{\pi} F(\theta,\phi) \cos(2m\theta) \cos(n\phi) d \theta d \phi, \\
b_{m,n} &= \frac{2}{\pi^{2}} \int_{0}^{2\pi} \int_{0}^{\pi}  F(\theta,\phi) \cos(2m\theta) \sin(n\phi) d \theta d \phi, \\
c_{m,n} &= \frac{2}{\pi^{2}} \int_{0}^{2\pi} \int_{0}^{\pi} F(\theta,\phi) \sin(2m\theta) \cos(n\phi) d \theta d \phi, \\
d_{m,n} &= \frac{2}{\pi^{2}} \int_{0}^{2\pi} \int_{0}^{\pi} F(\theta,\phi) \sin(2m\theta) \sin(n\phi) d \theta d \phi.
\end{align}
\end{subequations}
Note that 2$\theta$ is used to expand the distribution function $F(\theta,\phi)$ in Eq.~\eqref{eq12} because the domain of $\theta$ is from 0 to $\pi$, which is different from conventional 2D Fourier series expansion. Then, the arbitrary distribution function in the form of Fourier series is multiplied with the power densities of a dipole, and then is integrated over the polar angle and azimuth angle, as shown by Eq.~\eqref{eq10}. We find that the integral values of only four components are not zero among these integrations of the Fourier series, while the integral values of the rest components are all zero. Therefore, an arbitrary distribution function can be simplified to a distribution function with four Fourier coefficients, given by
\begin{equation}\label{eq15}
\mathcal{F}\left(\theta,\phi\right) =\frac{1}{2\pi^{2}}(1+2a\cos2\theta+2b\cos2\phi+4c\sin 2\theta \cos\phi+4d\cos2\theta \cos2\phi).
\end{equation}
with
\begin{subequations}\label{eq16}
\begin{align}
a &= \pi^{2} \lambda_{1,0} a_{1,0} = \int_{0}^{2\pi} \int_{0}^{\pi} F(\theta,\phi) \cos(2\theta) d \theta d \phi, \\
b &= \pi^{2} \lambda_{0,2} a_{0,2} = \int_{0}^{2\pi} \int_{0}^{\pi} F(\theta,\phi) \cos(2\phi) d \theta d \phi, \\
c &= \frac{\pi^{2}}{2} \lambda_{1,1} c_{1,1} = \int_{0}^{2\pi} \int_{0}^{\pi} F(\theta,\phi) \sin(2\theta) \cos(\phi) d \theta d \phi, \\
d &= \frac{\pi^{2}}{2} \lambda_{1,2} a_{1,2} = \int_{0}^{2\pi} \int_{0}^{\pi} F(\theta,\phi) \cos(2\theta) \cos(2\phi) d \theta d \phi.
\end{align}
\end{subequations}

Therefore, the power densities per solid angle of a dipole ensemble with arbitrary orientation distribution, for an emission angle $\alpha$ in half-space $j$, can be obtained by using Eq.~\eqref{eq10}:
\begin{subequations}\label{eq17}
\begin{align}
\label{eq17a}
\widetilde{P}^{s}_{j}\left(\alpha,a,b,c,d\right) &= \widetilde{P}^{s}_{j}(\alpha)\times\frac{1}{4}\left(1-a-b+d\right), \\
\widetilde{P}^{p1}_{j}\left(\alpha,a,b,c,d\right) &=
\widetilde{P}^{p1}_{j}(\alpha)\times\frac{1}{2}\left(1+a\right),\\
\label{eq17c}
\widetilde{P}^{p2}_{j}\left(\alpha,a,b,c,d\right) &=
\widetilde{P}^{p2}_{j}(\alpha)\times\frac{1}{4}\left(1-a+b-d\right),\\
\label{eq17d}
\widetilde{P}^{p3}_{j}\left(\alpha,a,b,c,d\right) &=
\widetilde{P}^{p3}_{j}(\alpha)\times c,
\end{align}
\end{subequations}
with
\begin{subequations}\label{eq18}
\begin{align}
\widetilde{P}^{s}_{j}(\alpha)&=\frac{k_{j}^{2}\cos\alpha}{\pi}\times\frac{3}{8}\frac{1}{k_{0}k_{z,0}}\frac{\left|1+a^{s}_{3-j}\right|^{2}}{\left|1-a^{s}\right|^{2}}T^{s}_{j}, \\
\widetilde{P}^{p1}_{j}(\alpha)&=\frac{k_{j}^{2}\cos\alpha}{\pi}\times\frac{3}{8}\frac{k^{2}_{\parallel}}{k^{3}_{0}k_{z,0}}\frac{\left|1+a^{p}_{3-j}\right|^{2}}{\left|1-a^{p}\right|^{2}}T^{p}_{j},\\
\widetilde{P}^{p2}_{j}(\alpha)&=\frac{k_{j}^{2}\cos\alpha}{\pi}\times\frac{3}{8}\frac{k_{z,0}}{k^{3}_{0}}\frac{\left|1-a^{p}_{3-j}\right|^{2}}{\left|1-a^{p}\right|^{2}}T^{p}_{j},\\
\widetilde{P}^{p3}_{j}(\alpha)&=-\frac{k_{j}^{2}\cos\alpha}{\pi}\times\frac{3}{8}\frac{k_{\parallel}}{k^{3}_{0}}\frac{1}{\left|1-a^{p}\right|^{2}}T^{p}_{3-j}T^{p}_{j}.
\end{align}
\end{subequations}
The effect of increasing coefficient $b$ on power density could also be achieved by decreasing coefficient $d$ as shown in Eq.~\eqref{eq17}, which indicates that Fourier coefficients $b$ and $d$ have the same effect on power density. Thus, only three distribution parameters are needed for description the relation between the dipole orientation and the power density. We define the distribution parameter $a$, $e$, and $c$, which are given by
\begin{subequations}\label{eq19}
\begin{align}
a &= a,\\
e &= \frac{b-d}{1-a},\\
c &= c.
\end{align}
\end{subequations}
Then, Eq.~\eqref{eq17} could be further simplified to
\begin{subequations}\label{eq20}
\begin{align}
\label{eq19a}
\widetilde{P}^{s}_{j}\left(\alpha,a,e,c\right) &= \widetilde{P}^{s}_{j}(\alpha)\times\frac{1}{4}\left(1-a\right)\left(1-e\right), \\
\widetilde{P}^{p1}_{j}\left(\alpha,a,e,c\right) &=
\widetilde{P}^{p1}_{j}(\alpha)\times\frac{1}{2}\left(1+a\right),\\
\label{eq19c}
\widetilde{P}^{p2}_{j}\left(\alpha,a,e,c\right) &=
\widetilde{P}^{p2}_{j}(\alpha)\times\frac{1}{4}\left(1-a\right)\left(1+e\right),\\ \label{eq19d}
\widetilde{P}^{p3}_{j}\left(\alpha,a,e,c\right) &=
\widetilde{P}^{p3}_{j}(\alpha)\times c,
\end{align}
\end{subequations}

\begin{figure}[!t]
\centering
\includegraphics[scale=1]{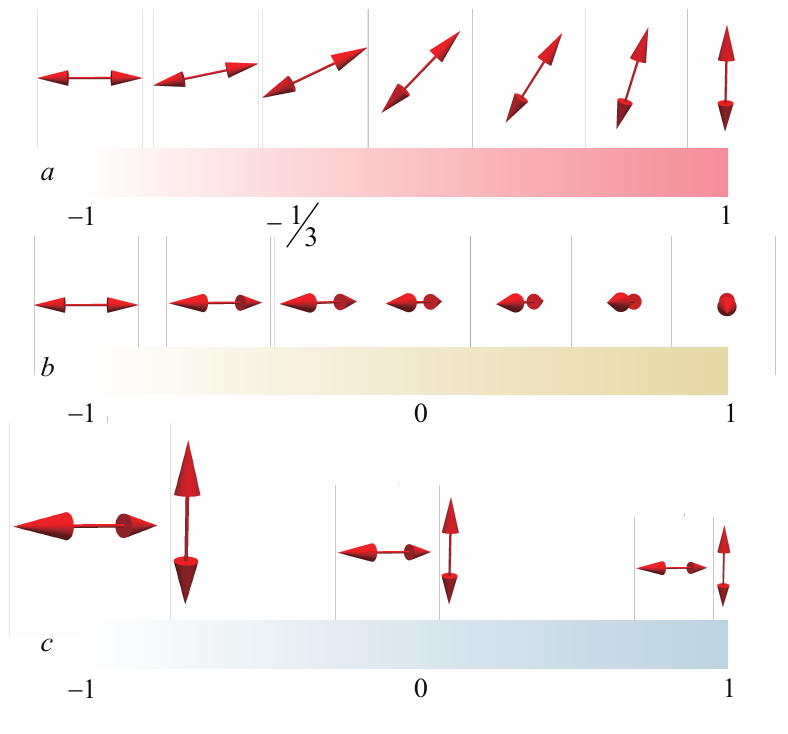}\\
\caption{\label{fig2}
Intuitive physical meaning of distribution parameters $a$, $e$, and $c$. The distribution parameters $a$ and $e$ describe the vertical and horizontal orientation distribution of the dipole ensemble respectively, while the distribution parameter $c$ describes the coupling between the vertical orientation distribution and the horizontal orientation distribution.}
\end{figure}

\section{Intuitive physical meaning}
In this section, specific distribution functions $F\left(\theta,\phi\right)$ are studied to reveal the intuitive physical meaning of the distribution parameters $a$, $e$, and $c$.

In the case that there is only one possible dipole orientation for the emitting dipoles in the OLED stacks represented by $(\theta_{1},\phi_{1})$ and its dipole vector is $\mathbf{d_{1}}$, the distribution function reads:
\begin{equation}\label{eq21}
\begin{split}
F\left(\theta,\phi\right) &= \delta(\mathbf{d}-\mathbf{d_{1}}) \times \sin \theta \\
 &=\delta\left(\theta-\theta_{1}\right)\delta\left(\phi-\phi_{1}\right),
\end{split}
\end{equation}
Considering Eq.~\eqref{eq16}, and Eq.~\eqref{eq19}, we can obtain the distribution parameters of a dipole:
\begin{subequations}\label{eq22}
\begin{align}
a &= \cos2\theta_{1}, \\
e &= \cos2\phi_{1}, \\
c &= \sin2\theta_{1}\cos\phi_{1}.
\end{align}
\end{subequations}
The distribution parameters $a$, $e$, and $c$ are controlled by two independent parameters $(\theta_{1},\phi_{1})$. As shown in Eq.~\eqref{eq22}, parameter $a$ is controlled by the factor $\cos2\theta_{1}$, where $\theta_{1}$ is the parameter describing the vertical orientation of the dipole, parameter $e$ is controlled by the factor $\cos2\phi_{1}$, where $\phi_{1}$ is the parameter describing the horizontal orientation of the dipole, and parameter $c$ is controlled by the factor $\sin2\theta_{1}\cos\phi_{1}$.

In the case that there are $n$ possible dipole orientations for the emitting dipole in the OLED stacks represented by $(\theta_{i},\phi_{i})$ and the corresponding dipole vectors are $\mathbf{d_{i}}$, the distribution function reads:
\begin{equation}\label{eq23}
\begin{split}
F\left(\theta,\phi\right)&=\frac{1}{n}\times\sum_{i=1}^{n}\left(\delta(\mathbf{d}-\mathbf{d_{i}})\sin\theta\right)\\
&=\frac{1}{n}\times\sum_{i=1}^{n}\left(\delta\left(\theta-\theta_{i}\right)\delta\left(\phi-\phi_{i}\right)\right),
\end{split}
\end{equation}
The same procedure is easily adapted to obtain the distribution parameters of $n$ possible dipole orientations:
\begin{subequations}\label{eq24}
\begin{align}
a &= \frac{1}{n}\sum_{i=1}^{n}\left(\cos2\theta_{i}\right), \\
e &=
\frac{\sum_{i=1}^{n}((1-\cos2\theta_{i})\cos2\phi_{i})}{\sum_{i=1}^{n}(1-\cos2\theta_{i})}, \\
c &= \frac{1}{n}\sum_{i=1}^{n}\left(\sin2\theta_{i}\cos\phi_{i}\right).
\end{align}
\end{subequations}
When there are $n$ possible dipole orientations in the OLED stacks, distribution parameters $a$, $e$, and $c$ are controlled by $2n$ independent parameters $(\theta_{i},\phi_{i})$. As shown by Eq.~\eqref{eq24}, parameter $a$ is controlled by the factor $\cos2\theta_{i}$, where $\theta_{i}$ is the parameter describing the vertical orientation of $n$ dipoles, parameter $e$ is controlled by the factor $\cos2\phi_{i}$, where $\phi_{i}$ is the parameter describing the horizontal orientation of $n$ dipoles, and parameter $c$ is controlled by the factor $\sin2\theta_{i}\cos\phi_{i}$.

Therefore, parameters $a$ and $e$ describe the vertical and horizontal orientation distribution of a dipole ensemble respectively, and parameter $c$ describes the coupling between the vertical and horizontal distribution, as illustrated in Fig.~\ref{fig2}(a). The parameter $a$ describes the vertical distribution of a dipole ensemble. When $-1 \leq a < -1/3$, the orientation of a dipole ensemble tends to parallel to the horizontal $x-y$ plane; when $a = -1/3$, the orientation of a dipole ensemble is randomly distributed; when $-1/3 < a \leq 1$, the orientation of a dipole ensemble tends to parallel to $z$ axis. The parameter $e$ describes the horizontal distribution of the dipole ensemble. When $-1 \leq e < 0$, the orientation of the dipole ensemble tends to parallel to the $y$ axis, when $e = 0$, the orientation of the dipole ensemble is randomly distributed; when $0 < e \leq 1$, the orientation of the dipole ensemble tends to parallel to the $x$ axis. Note that when $a=1$, the denominator of $e$ is equal to zero, which is ill defined but consistent with the physical meaning since there are only vertical dipoles when $a=1$. In that case, we could define $e=0$ when $a=1$, which also represents random horizontal distribution. The parameter $c$ describes the coupling. When $c<0$, the coupling between the vertical and horizontal distribution is positive for the power densities increase as shown by Eq.~\eqref{eq17d}; when $c=0$, there is no coupling and dipoles can be decomposed into vertical dipole and horizontal dipole which is wildly applied in the previous theories; when $c>0$, the coupling effect is negative. For the distribution parameter $a$, it is related to $\Theta$ and $S$~\cite{Schm2017Emit}, previously reported to describe the vertical orientation distribution of dipoles, given by
\begin{equation}\label{eq25}
a=2\Theta-1=\frac{1}{3}\left(4S-1\right).
\end{equation}

Note that other definition of three distribution parameters to simplify Eq.~\eqref{eq17} is allowed, which will lead to different physical meaning of the distribution parameters. For example, we could define the distribution parameters $d_{x}$, $d_{z}$, and $d_{x,z}$, given by
\begin{subequations}\label{eq26}
\begin{align}
d_{x} &= \frac{1}{4}(1-a+b-d),\\
d_{z} &= \frac{1}{2}(1+a),\\
d_{x,z} &= c.
\end{align}
\end{subequations}
Then, Eq.~\eqref{eq17} could be further simplified to
\begin{subequations}\label{eq27}
\begin{align}
\label{eq21a}
\widetilde{P}^{s}_{j}\left(\alpha,d_{x},d_{z},d_{x,z}\right) &= \widetilde{P}^{s}_{j}(\alpha)\times(1-d_{x}-d_{z}), \\
\widetilde{P}^{p1}_{j}\left(\alpha,d_{x},d_{z},d_{x,z}\right) &=
\widetilde{P}^{p1}_{j}(\alpha)\times d_{z},\\
\label{eq21c}
\widetilde{P}^{p2}_{j}\left(\alpha,d_{x},d_{z},d_{x,z}\right) &=
\widetilde{P}^{p2}_{j}(\alpha)\times d_{x},\\
\label{eq21d}
\widetilde{P}^{p3}_{j}\left(\alpha,d_{x},d_{z},d_{x,z}\right) &=
\widetilde{P}^{p3}_{j}(\alpha)\times d_{x,z}.
\end{align}
\end{subequations}
Under this definition, distribution parameter $d_{x}$ represents the ratio of $x$ component of the emitting dipoles, distribution parameter $d_{z}$ represents the ratio of $z$ component of the emitting dipoles, and distribution parameter $d_{x,z}$ represents the interference of the $x$ and $z$ component of the emitting dipoles caused by the uneven distribution of $x$ and $z$ component of the emitting dipoles, which is also the meaning of the coupling between the horizontal and vertical distribution of parameter $c$.

\begin{figure*}[!t]
\centering
\includegraphics[scale=1]{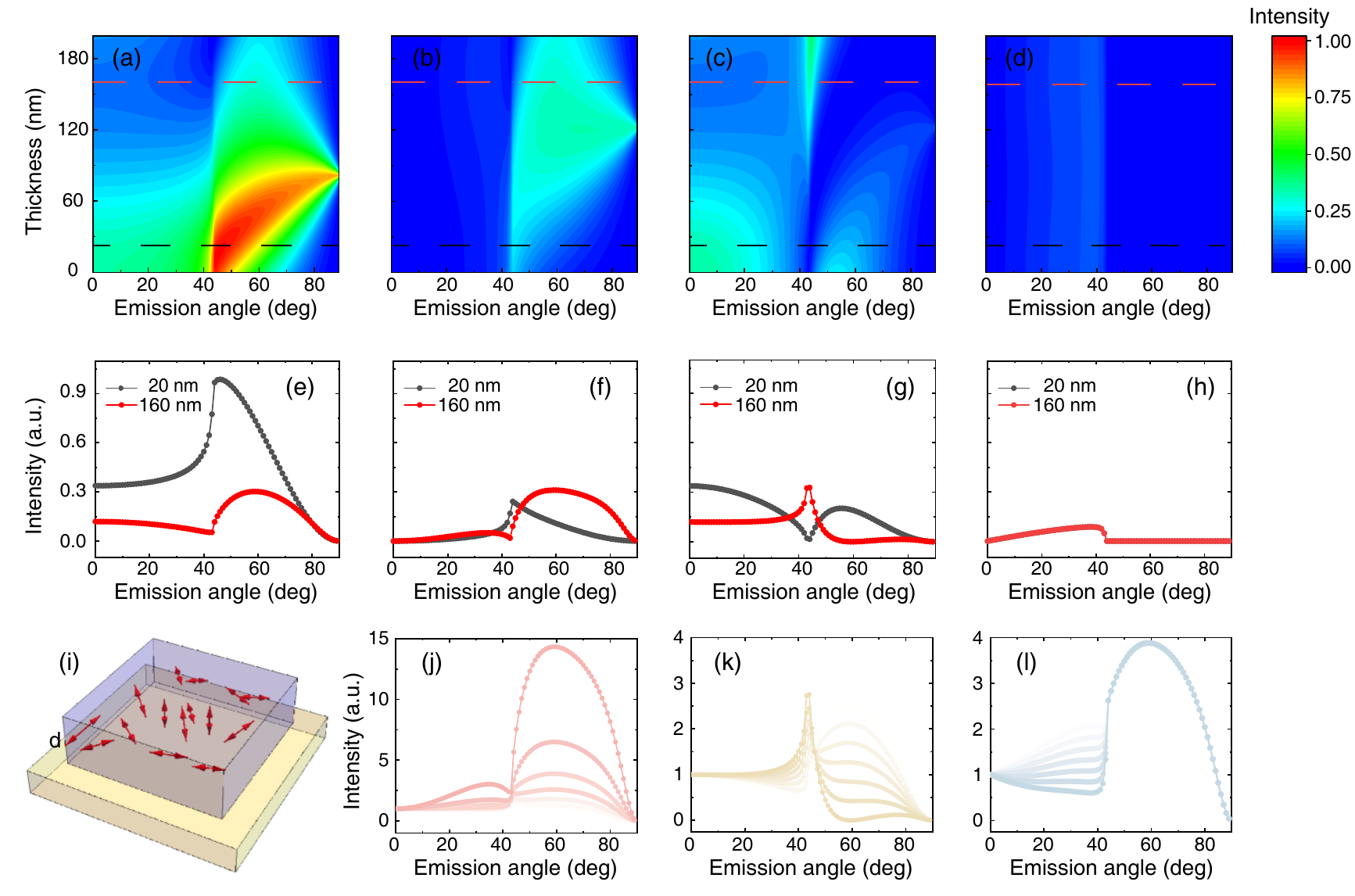}\\
\caption{\label{fig3}
(a)-(d) Simulated different energy densities $\widetilde{P}^{s}_{j}\left(\alpha,a,e,c\right)$, $\widetilde{P}^{p1}_{j}\left(\alpha,a,e,c\right)$, $\widetilde{P}^{p2}_{j}\left(\alpha,a,e,c\right)$, and $\widetilde{P}^{p3}_{j}\left(\alpha,a,e,c\right)$ at different emission angle and different thickness. (e)-(h) The change of power density with emission angle at 20 nm and 160 nm. (i) Sample for measuring the orientation distribution parameters. The refractive indices of glass substrate and organic thin film are 1.524 and 1.6, and the simulated wavelength is 400 nm (j) Simulated signals with different parameter $a$. (k) Simulated signals with different parameter $e$. (l) Simulated signals with different parameter $c$.}
\end{figure*}

\section{Proposed experiment configuration}
To determine the orientation distribution of the emitting dipole ensemble in the emitting layer~\cite{doi:10.1063/1.4907890,doi:10.1021/cm500802p}, the test structure we considered is shown in Fig.~\ref{fig3}(i). An organic thin film is evaporated on the glass substrate, whose thickness is $d_{0}$. The emitting dipoles are doped in the middle of the emitting layer. The refractive indices of glass substrate and organic thin film are 1.524 and 1.6, respectively, and the simulated wavelength is 400 nm. By measuring the intensity of the spectrum along the glass substrate side at different angles, we can get the intensity changing with respect to the emission angle~\cite{Has2018Well}. Then the intensity of the spectra at 0 emission angle is normalized to 1. These normalized spectra usually have different shapes, which indicate different orientation distribution of the emitting dipoles~\cite{Chris2008Intrinsic}.

Since there was only one distribution parameter $\Theta$ or $S$ in the previous description of emitting dipole orientation, only $p$-polarized spectra were considered in previous experiment configuration to measure the orientation distribution of the emitting dipole ensemble~\cite{Fris2010Deter,Kom2017Dip}. However, in our proposed experiment configuration, non-polarized spectra should be used for analysis, in order to contain all information of these three distribution parameters. The normalized non-polarized spectra consist of four components $\widetilde{P}^{s}_{j}(\alpha, a,e,c)$, $\widetilde{P}^{p1}_{j}(\alpha, a,e,c)$, $\widetilde{P}^{p2}_{j}(\alpha, a,e,c)$, and $\widetilde{P}^{p3}_{j}(\alpha, a,e,c)$, and they are adjusted by three independent parameters $a$, $e$, and $c$, as shown by Eq.~\eqref{eq20}. If only normalized $p$-polarized spectra are used for analysis, the distribution parameter $e$ cannot be obtained.

In our proposed experiment configuration, the use of non-polarized spectra requires us to consider $s$-polarized spectra on the basis of the previous analysis of $p$-polarized spectra. In order to distinguish the $s$-polarized power density and $p$-polarized power densities of different shapes, the thickness of the emitting layer should be optimized, as illustrated in Fig.~\ref{fig3}(a)-(h). The $s$-polarized and $p$-polarized power densities $\widetilde{P}^{s}_{j}(\alpha, a,e,c)$, $\widetilde{P}^{p1}_{j}(\alpha, a,e,c)$, $\widetilde{P}^{p2}_{j}(\alpha, a,e,c)$, and $\widetilde{P}^{p3}_{j}(\alpha, a,e,c)$ are supposed to be in different shapes but in the same order of magnitude. When the thickness of the emitting layer approaches 0, the total $s$-polarized power density is much stronger than the total $p$-polarized power densities, where the total power density is a sum of power densities at every emission angle. The $p$-polarized power densities have little effect on the total power density, and the dipole orientation distribution information contained in the $p$-polarized power densities is easily concealed by the noise during the measurement. By increasing the thickness of the emitting layer, as shown in Fig.~\ref{fig3}(a), the $s$-polarized power density begins to decrease, and the $p$-polarized power densities begin to increase. The optimal thickness of the emitting layer is near 160 nm, where all four power densities have a significant impact. Finally, we design an optimal experiment configuration in which three distribution parameters could be measured precisely. When the thickness of the emitting layer is 160 nm, non-polarized spectra corresponding to different distribution parameters are shown in Fig.~\ref{fig3}(j)-(l).

For an electric dipole, it hardly radiates along the direction of its dipole vector. As the distribution parameter $a$ approaches 1, the orientation of the dipole vector tends to parallel to the $z$ axis, and its intensity at 0 emission angle becomes weak. Thus, the normalized non-polarized spectra have stronger intensities at large emission angles, as illustrated in Fig.~\ref{fig3}(j). This intuitive property of an electric dipole also leads to the anisotropy of its intensity when the horizontal orientation distribution of the dipole is anisotropic. When the orientation of dipole vector parallels to the $x$ axis, its intensities in $x-z$ plane at large emission angles become weak. Therefore intensities in $x-z$ plane at large emission angles become weaker as the distribution parameter $e$ approaches 1, shown in Fig.~\ref{fig3}(k). The coupling between horizontal and vertical components would have effect on the power density of the OLED devices only if the light could radiated towards both upward and downward direction of the OLED devices. When the emission angle becomes larger, total internal reflection occurs at the interface between the air layer and the emitting layer and the distribution parameter $c$, describing the coupling effect, has no effect on the power density. Therefore, the distribution parameter $c$ only affects intensities where the emission angles are smaller than the total internal reflection angle, as illustrated in Fig.~\ref{fig3}(l).

\section{Summary and conclusions}

By introducing an arbitrary distribution function to expand a dipole into a dipole ensemble, we rigorously prove that only three distribution parameters are needed to fully describe the relationship between the orientation distribution of a dipole ensemble and its power density. These three distribution parameters could adjust the ratio of different parts of the power density, which means that we can determine these three distribution parameters by measuring the power intensity radiated by the dipole ensemble in OLED stacks.

We discuss an optimal experiment configuration for measuring the three distribution parameters and intuitively explain the reason why these three distribution parameters $a$, $e$, and $c$ could be obtained from the normalized non-polarized spectrum. With the improvement of OLED manufacturing technology and more precise control of the orientation of dipoles in OLED stacks, our method fully describes the orientation distribution of dipoles in OLED stacks, thereby providing guidance for OLED manufacturing technology.

\section{Acknowledgments}
We thank Xiaoyuan Hou and Shaobo Liu for useful discussions about experiment configuration, and Zhenghong Li for fruitful discussions. This work was supported by the China National Key Basic Research Program ( 2016YFA0301103, 2016YFA0302000 and 2018YFA0306201) and the National Science Foundation of China (11774063, 11727811, 91750102 and 91963212). L.S. was further supported by the Science and Technology Commission of Shanghai Municipality (19XD143600, 2019SHZDZX01 and 19DZ2253000).

\section{Disclosures}
Haiwei Yin have financial interest in Ideaoptics Instruments Co., Ltd. The remaining authors declare that they have no conflict of interest.

\section{Data Availability Statement}
No data were generated or analyzed in the presented research.

\bibliography{paper.bib}
\end{document}